# CARS and SHG microscopy to follow the collagen production in living human corneal fibroblasts and mesenchymal stem cells in fibrin gel 3D cultures

L. Mortati*, C. Divieto and M.P. Sassi

INRIM - Istituto Nazionale di Ricerca Metrologica
Strada delle cacce, 91 – 10135 Torino (Italy)
*l.mortati@inrim.it

**ABSTRACT**

Coherent anti-Stokes Raman scattering (CARS) microscopy is combined with second harmonic generation (SHG) technique in order to follow the early stage of stem cell differentiation within a 3D scaffold.

One of the first evidence of hMSCs differentiation is the formation of an extracellular matrix (ECM) where the collagen protein is its main component.

This work demonstrated the multimodal CARS and SHG microscopy as a powerful non-invasive label free technique to follow the collagen production in living cell 3D cultures. Its ability to image the cell morphology and the produced collagen distribution on the same sample at the same time, on a long term (4 weeks) experiment allowed to obtain important information about the cell-scaffold interaction and the ECM production. The very low limit reached in detecting collagen has permitted to map even the small amount of collagen produced by the cells in few hours of culture. This demonstrates multimodal CARS and SHG microscopy as a novel method to follow cells collagen production and cells differentiation process in both short and long term experiments. In addition the experiment shows that the technique is a powerful tool for imaging of very thick sections (about 4 mm) with several advantages in its applications. As collagen production is considered a biomarker for ECM production and also a signal of initial stem cells differentiation, the study conducted on mesenchymal stem cell in 3D cultures confirmed that differentiation stimulus is induced by the fibrin gel scaffold.

Coherent anti-Stokes Raman scattering (CARS) and second harmonic generation (SHG) microscopy techniques, allow a label free non-invasive chemical selection of target species, deep tissue penetration and high 3D spatial resolution at a cellular level. CARS microscopy can detect lipid membranes and droplet compartments in living cells and SHG microscopy enables a strong imaging contrast for molecules with a non-centrosymmetric ordered structure like collagen. In addition, CARS and SHG techniques can be easily combined together in the same microscopy allowing multiple chemical contrasts.

This study aims to demonstrate the efficacy of the CARS-SHG combined technique to investigate, in a non-invasive and non-destructive experiment, the dynamics and the distribution of the collagen produced by living stem cells seeded in a 3D fibrin scaffold. The monitoring of stem cell differentiation within a scaffold in a non-destructive way will be an important advantage in regenerative medicine and tissue engineering field.

**INTRODUCTION**

Regenerative medicine uses biomimetic materials (scaffolds) and cells to facilitate regeneration of tissues and organs *in vivo* [1, 2, 3].

Mesenchymal stem cells (MSCs) have been widely studied in the last twenty years in regenerative medicine applications [3] because of their role to maintain the cell natural turnover, replacing differentiated cells naturally expired or damaged/dead due to an injury or a disease [4, 5]. MSCs can be isolated from the bone marrow aspirates or from other sources such as adipose tissue, umbilical cord blood and placenta [6, 7]. They can be cultured *in vitro* and can be induced to differentiate in several cell types, such as bone, cartilage, tendon, cardiac muscle, skeletal muscle, neural cells, adipose and connective tissue [3, 4, 8, 9].

The use of biomimetic scaffold offers to MSCs the native physiologic-like and three-dimensional environment which surrounds *in vivo* tissues and organs [3]. Recent studies show that scaffold chemical composition, internal architecture and stiffness are parameters of influence for the behaviour and functions of the cells seeded on it [10-12]. This makes extremely important that interactions between scaffold and stem cells should be studied *in vitro* before implanting the cell-scaffold construct.

In their native environment, stem cells are surrounded by a 3D extracellular matrix (ECM) produced by fibroblasts. ECM is involved in regulating cell behaviour (e.g. survival and proliferation) and function (e.g. differentiation) and it is composed by several proteins and other macromolecules. Collagen is the main ECM component and the most abundant protein in mammals and it gives a mesh structure to the ECM [13, 14]. In stem cell biology, it is well known that MSCs cultured *in vitro* can differentiate producing ECM when chemically or mechanically stimulated [9, 15,16,17].

Collagen production represents one of the first steps of ECM formation [14] and it is normally assessed *in vitro* as indicator of cell differentiation process [18, 19, 20]. Collagen has a non-centrosymmetric ordered triple-helix structure with a very high level of crystallinity, that is a suitable condition for enabling second harmonic generation process [21].

 The dynamics of collagen production can be studied as biomarker of cell differentiation with several techniques. However, conventional techniques are destructive and/or invasive and/or do not provide information about spatial distribution of the protein. Gene expression analysis contemplates destructive procedure steps and loses the spatial distribution of the protein. Protein analysis with western blot needs to destroy samples and requires a consistent amount of protein. Protein analysis with immunofluorescence techniques gives information on the spatial (2D) distribution of the protein but samples need to be fixed and/or sectioned. With traditional techniques it is then impossible to follow, in non-destructively short and long term experiments, the dynamic distribution of collagen produced from the same sample. A non-destructive and non-invasive procedure based on CARS and SHG combined microscopy, able to investigate the cell-scaffold interaction in term of cell behaviour and functions over the time is discussed in this work.

In this work the collagen produced by living human corneal fibroblasts (hCFs) and human mesenchymal stem cells (hMSCs) seeded in a fibrin gel matrix has been monitored over the time. Fibrin gel is largely used in regenerative medicine as biodegradable, biocompatible and non-cytotoxic scaffold, capable to induce both osteogenic and chondrogenic stem cell differentiation [22, 23]. Fibrin gel scaffold supports and stimulates through its composition and stiffness, in absence of any external chemical and mechanical stimulus, stem cells differentiation with collagen production [24]. hCFs have been chosen as positive control for collagen production because they are normally devoted to produce collagen *in vivo* and they maintain this property also when cultured *in vitro* [5, 25, 26] . hMSCs have been chosen to investigate the fibrin gel capability to induce stem cells collagen production without any other kind of stimuli.

Multiphoton microscopy based on the combination of multiple nonlinear optical phenomena like CARS and SHG [27], allows deep tissue penetration [28] and high 3D spatial resolution [29], that are powerful requirements in analyzing thick tissue sections at cellular level. CARS microscopy provides chemical contrast from Raman-active molecular vibrations and it is able to detect membranes and lipid droplet compartments in living cells, like fibroblasts [30, 31] tuning the excitation sources to the $CH_2$ symmetric stretching vibration wavenumber at around 2845 $cm^{-1}$.

SHG microscopy is based on a second order nonlinear optical process and it is sensitive to the molecular structures, enabling a strong imaging contrast for non-centrosymmetric molecular ordered structures such as collagen [32-37], microtubule arrays [38], skeletal muscle myosin [39, 40], cell membranes [41], and also cellulose [42]. Recently it has been demonstrated that CARS and SHG techniques can be easily combined together in the same microscopy allowing multiple chemical contrasts [43-46].

The aim of this work is to demonstrate that a combined CARS and SHG microscopy is an adequate and optimal technique to follow on different time scales the collagen produced by cells cultured in a 3D scaffold, with a non-invasive, non-destructive and label free method.

**EXPERIMENTAL**

**CARS and SHG Microscopy**

A passively mode-locked Nd:YVO$_4$ (Yttrium Vanadate crystal doped with Neodymium) laser emitting at 1064 nm (Picotrain, HighQLaser) was used as a master source for CARS and SHG microscopy. This optical source emits a continuous train of 10 ps pulses at the repetition rate of 76 MHz and it is equipped with a SHG unit. The 532 nm output of the SHG frequency doubling has a pulse width of about 5 ps and it is used to synchronously pump an Optical Parametric Oscillator (OPO) (Levante Emerald, APE Berlin). The OPO acts as the tuneable source and it is based on a non linear Lithium-Triborate (LBO, LiB$_3$O$_5$) crystal as a parametric amplifier in a resonant optical cavity. The tuning range is from 700 nm to 1020 nm for the signal wave and from 1110 nm to 2200 nm for the idler wave. Signal and idler beams exit collinear in our experiments, entering the scanning unit (FluoView FV300, Olympus) combined with the upright microscope (BX51WI, Olympus). This allows a punctual detection of CARS and SHG signals all over the sample with high resolution and high excitation efficiency. The Z depth scanning is achieved by moving the focusing objective with a stepping motor.

In order to focus the excitation beams on living cells samples, a water immersion objective (LUMPLFLN 60XW NA=1 W.D.=2 mm, Olympus) fully compensate for both spherical and chromatic aberrations from the UV to the near infrared region was used. The water immersion objective was cleaned and sterilized with a solution 70% ethanol in water (v/v) before each imaging experiment.

The forward de-scanned CARS and SHG signals are collected through an objective (UPLSAPO 20x objective NA=0.75 W.D.=0.6 mm, Olympus) and focused on a PMT (R3896, Hamamatsu) with a plano-convex lens with a focal length 25 mm.

Cell membranes and rich lipidic structures were imaged using CARS process looking at the $CH_2$ symmetric stretch Raman modes at around 2844 $cm^{-1}$. The CARS signal was generated at around 731.8 nm by tuning the pump and the Stokes beams to 924.1 nm and 1253.7 nm respectively.

Collagen structures were imaged using SHG process, tuning the OPO signal to 950 nm and detecting the corresponding SHG wavelength at 475 nm.

In order to further block the residual excitation beams and transmit the CARS and the SHG signals respectively, bandpass filters centered at 716 nm with 43 nm bandwidth (FF01-716/43, Semrock) and at 480 nm with 20 nm bandwidth (BP470-490, Chroma Technology), both coupled with shortpass filters with 770 nm of cut-off wavelength (FF01-770/SP, Semrock) are placed before the detector.

In order to prevent sample damages and optimize the output signal the excitation beams were attenuated through a neutral density variable filter wheel (NDC-50C-4M, Thorlabs).

CARS and SHG imaging are obtained in sequence on the same sample and SHG images were processed using ImageJ software in order to enhance their contrast.

**Cell selection and culture in fibrin gel scaffolds**

Human mesenchymal stem cells (hMSCs) were purchased by Lonza (Basel, Switzerland). They are bone marrow derived-hMSCs from donor. hMSCs were expanded and maintained in a complete non differentiating growth medium (MSCBM, Lonza) supplemented with 10% fetal bovine serum (FBS), 2% L-glutamine, 0,1% antibiotics (gentamicin and amphotericin B) (Lonza).

Human corneal fibroblasts (hCF) were kindly provided by Dr Sizzano (Transplants Centre of the Regione Piemonte). hCFs were expanded, maintained and resuspended in DMEM (Listarfish, Milano, Italy) containing 10% FBS, 1% L-Glutamine and 1% kanamycin. Cells were cultured until they reached 80-85% of confluency (cells surface per total area), then they were washed with 1X Phosphate Buffer Saline (PBS), detached with 0.05% trypsin/0.53 mM EDTA, counted by means of a hemocytometer and suspended at $1x10^6$ cells per 200 μL complete growth medium to be seeded within the scaffolds of fibrin gel. hMSCs were used at passage 8 and hCFs at passage 10 to prepare the fibrin gel and cells constructs.

**Scaffolds of fibrin gel preparation**

Fibrin gel scaffolds were prepared from fibrinogen and thrombin, both proteins involved in blood clotting. Fibrinogen (5 mg/mL) (Sigma, USA; Cod.F8630) was reconstituted in PBS 1x and thrombin (25 ug/mL) (Sigma, USA; Cod. T9549) was reconstituted in PBS 1x.

3D fibrin scaffolds containing cells were prepared in 35x10 mm cell culture dishes (CytoOne) by mixing in the following order: 325 μL of thrombin (25 ug/mL), 125 μL of medium, 200 μL of cell suspension, 1350 μL of fibrinogen (5 mg/mL). The components were allowed to polymerize undisturbed at room temperature for 5 minutes in order to obtain fibrin 3D scaffolds with thickness of about 4 mm. Then the scaffolds were covered with 2 mL of cell culture medium and placed in an incubator for cell cultures under controlled conditions of temperature (37°C) and $CO_2$ (5%). Cells within the scaffolds were fed every 3–4 days by completely replacing the medium with fresh medium. Five samples were prepared for both cell types, in order to detect the collagen production at days 0, 7, 14, 21, 28.

**RESULTS AND DISCUSSION**

Imaging of living cells (hCFs and hMSCs) and collagen was done using CARS and SHG microscopy technique in a time-course experiment at different days in culture (day 0, 7, 14, 21, 28). Cells were localized using CARS microscopy technique looking for cell rich lipidic structures and imaging was done in three dimensions with a step in the Z-axis of about 800 nm. The overall Z scan range was chosen accordingly to the 3D extension of the cells in the fibrin gel. The XY pixel pitch was equal to 0.230178 μm/pixel for all the measurements, while the image size was adapted according to cells shape in each experiment.

The overall 3D imaging lasted between 5 and 10 minutes according to the Z scan range and for each Z-step up to nine images were acquired and adaptively averaged using a Kalman filter.

Pixel dwell time was around 10 μs and the average power at the sample was about 25 mW for the pump signal and less than 10 mW for Stokes signal.

The slices obtained for each experiment were used to create a maximum intensity Z-projection image, in order to have in a single picture all the interesting extension of the cell in the measured volume.

Collagen detection was performed using SHG microscopy technique after CARS imaging, keeping the same dimensional and temporal parameters of the related CARS imaging. The average excitation power at the sample was about 20 mW.

A maximum intensity Z-projection was done also for SHG imaged slices, obtaining a single picture of the measured volume. Since the SHG image contrast was weak, Z-projected images (Fig. 1A) were processed using ImageJ software contrast enhancement (Fig. 1B) followed by a Gaussian blurring with one pixel size (Fig. 1C) and a manual brightness/contrast adjustment (Fig. 1D).

After these operations, the SHG intensity image became a quasi-binary image, holding all the information related to the collagen distribution in a clearer manner.

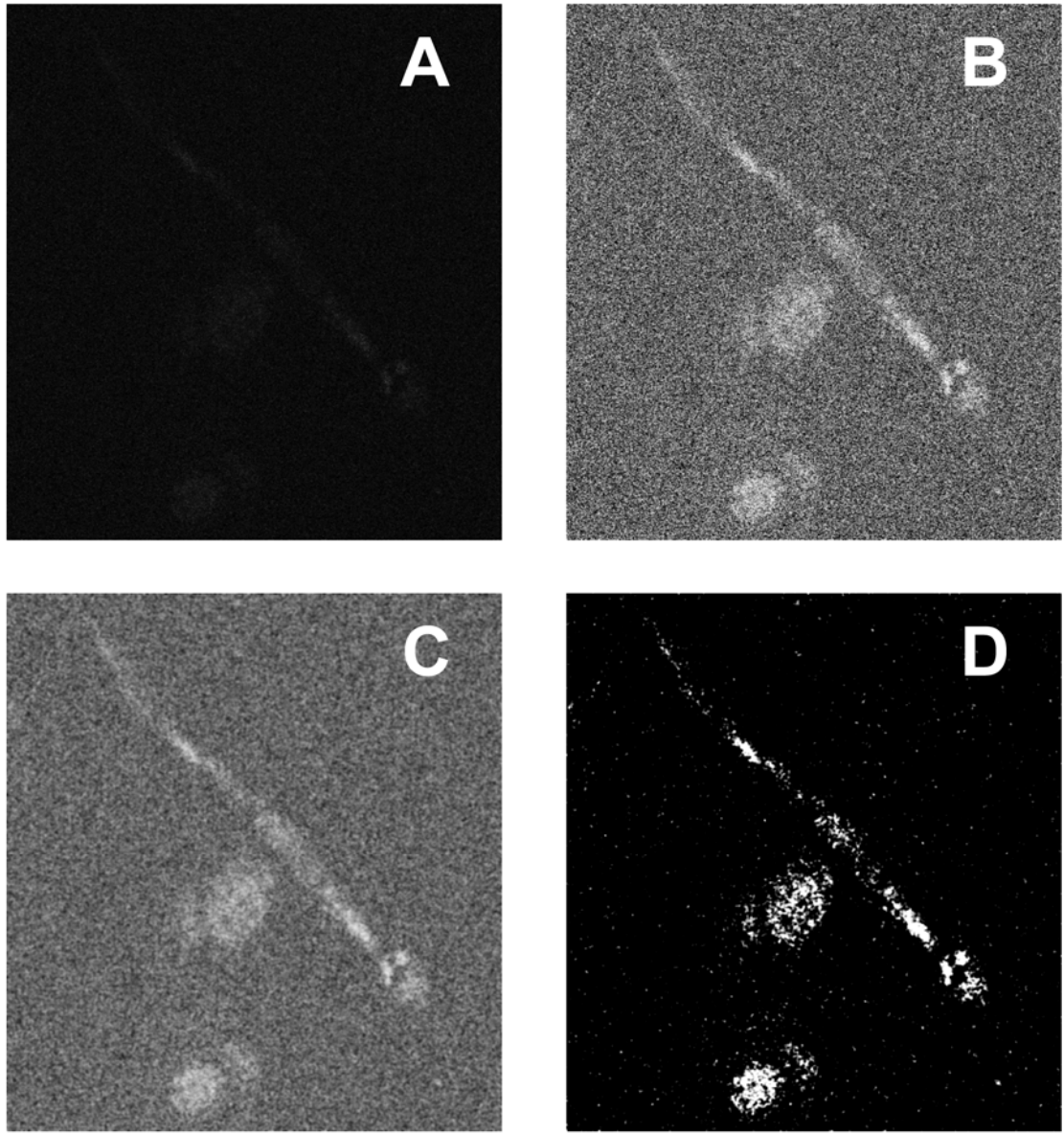

**Fig. 1 - A sequence of adjustments have been performed to improve SHG image contrast obtaining quasi-binary images with a clearer visualization of the collagen produced by the cells.**

Membranes and lipid droplets inside the cells were clearly visible using CARS microscopy tuned at 2844 cm$^{-1}$ and collagen was detected using SHG microscopy.

Cells to be imaged have been chosen selecting those being in areas of the fibrin matrix with a mid-level of confluency.

At culture day 0 both hCFs (Fig. 2A) and hMSCs (Fig. 2B) show irrelevant SHG signal arising from collagen as expected since the cells need some time to produce collagen and ECM. Nevertheless hCFs show a slight initial collagen formation, indicating that the technique has a very low limit of detection and it is able to detect collagen production after only few hours of culture.

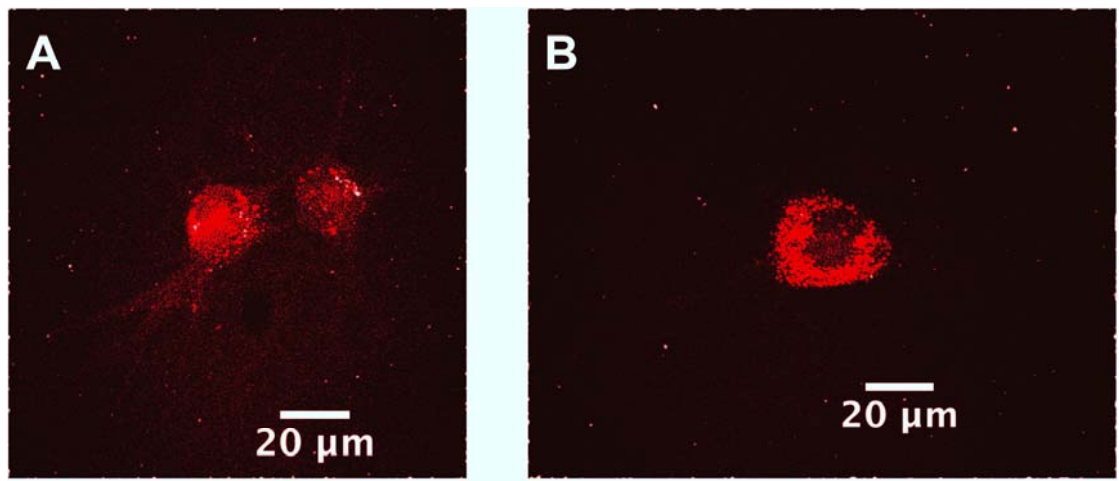

**Fig. 2 - hCFs (A) and hMSCs (B) (both in red) at day 0 showed only a slight or no collagen production (in white).**

This also indicates that for the used SHG detection scheme the eventual two-photon autofluorescence generated from natively fluorescent compounds inside the cells (like NAD(P)H or oxidized flavoproteins) is not an issue. The spurious white spots in the images outside the cells come from SHG signal enhancement process noise.

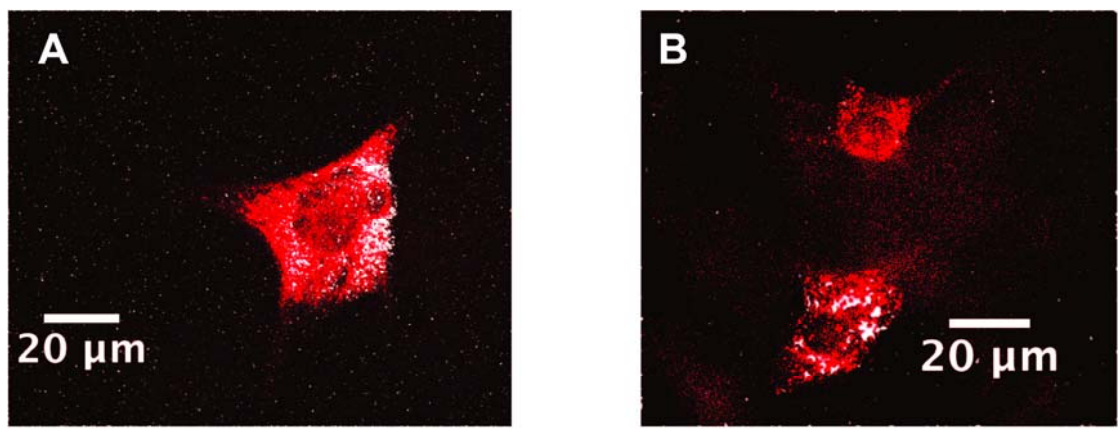

**Fig. 3 - hCFs (A) and hMSCs (B) (in red) at day 7 showed collagen production (in white).**

At day 7 both hCFs (Fig. 3A) and hMSCs (Fig. 3B) show relevant SHG signal arising from collagen produced by the cells. In hCFs sample collagen produced by the cells is clearly localized. In hMSCs sample some cells begin to produce collagen indicating an initial differentiation process with the formation of the ECM. This demonstrates that SHG microscopy can detect small amounts of collagen produced in a short period of time by living cells without fixing and staining the sample.

Collagen production observed in hMSCs sample demonstrates the ability of this technique to discriminate the initial differentiation step in living stem cells. The results obtained demonstrate that CARS and SHG microscopy is able to investigate non-invasively the stem cells differentiation induced by the scaffold and confirm that fibrin scaffold properties can induce stem cells differentiation.

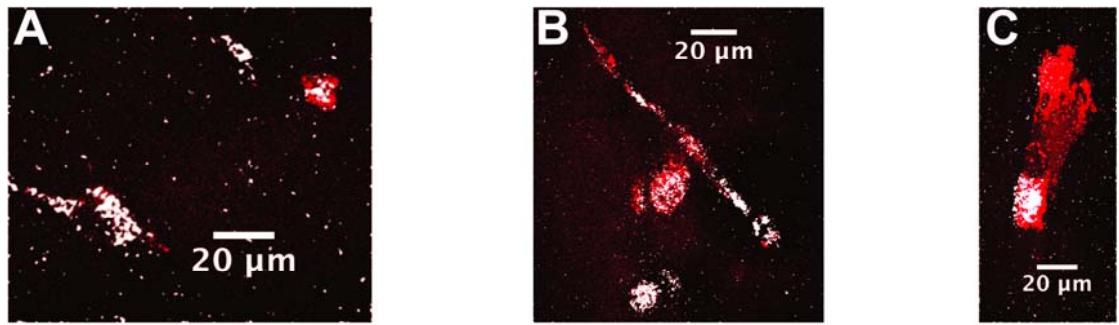

Fig. 4 - hCFs at days 14 (A), 21 (B), 28 (C) (in red) showed collagen production (in white).

Collagen was detected also at days 14, 21 and 28 for both hCFs (Fig 4A-4C) and hMSCs (Fig. 5A-5C).

In this experiment variations on the distribution of collagen produced by cells at different culture days is not biologically relevant, because images acquired on different analyzed samples, are not related to the same cell or sample region. Cells in the fibrin gel are living and proliferating and continue to duplicate in culture even if submitted to subsequent measurements. This is a further proof that laser power used for imaging is safe and does not induce any relevant influence on cell growth.

It is important to stress that, as it is shown in Fig. 5A-5C, hMSCs produced collagen also at days 14, 21 and 28, confirming that fibrin gel scaffold induced hMSCs to produce collagen and hence ECM.

In the majority of the acquired images, collagen produced by hCFs and hMSCs is not localized in correspondence of the nuclear region (the darker round shape region generally situated in the center of the cells). This is a further confirmation that the SHG signal is generated by the collagen protein, because proteins are produced in the cytoplasm of the cell and not in the nucleus.

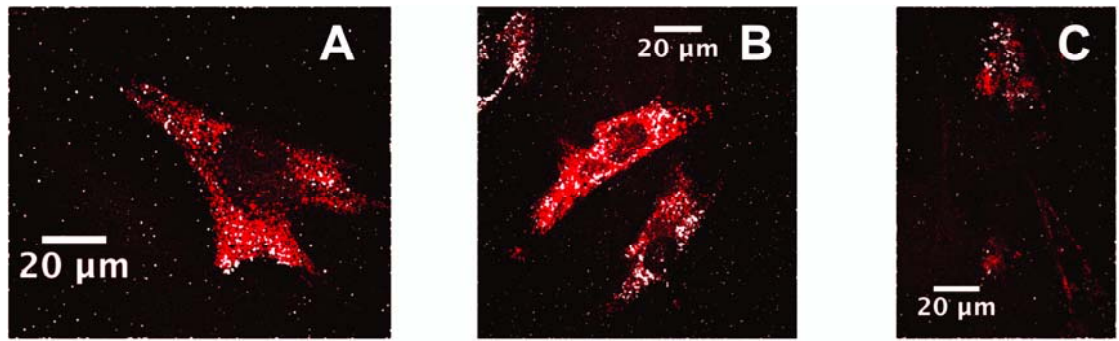

**Fig. 5 - hMSCs at days 14 (A), 21 (B), 28 (C) (in red) showed collagen production (in white).**

**CONCLUSION**

This work demonstrated the multimodal CARS and SHG microscopy as a powerful non-invasive label free technique to follow the collagen production in living cell 3D cultures. Its ability to image the cell morphology and the produced collagen distribution on the same sample at the same time, on a long term (4 weeks) experiment allowed to obtain important information about the cell-scaffold interaction and the ECM production. The very low limit reached in detecting collagen has permitted to map even the small amount of collagen produced by the cells in few hours of culture. This demonstrates multimodal CARS and SHG microscopy as a novel method to follow cells collagen production and cells differentiation process in both short and long term experiments. In addition the experiment shows that the technique is a powerful tool for imaging of very thick sections (about 4 mm) with several advantages in its applications. As collagen production is considered a biomarker for ECM production and also a signal of initial stem cells differentiation, the study conducted on mesenchymal stem cell in 3D cultures confirmed that differentiation stimulus is induced by the fibrin gel scaffold.

The results of this work open new perspectives for tissue engineering and regenerative medicine to investigate rapidly and in a non-invasive way living cell cultures, enabling a better understanding of the interaction between cells and scaffolds and to detect early ECM formation, in the initial stages of the stem cells fibroblastic, chondrogenic and osteogenic differentiation processes. This technique offers advantages to characterize and to evaluate the performance of the cell-ECM/scaffold constructs before implanting them into the donor in pre-clinical and clinical applications of regenerative medicine.

**AKCNOLEDGMENTS**

The authors like to thanks Dr Sizzano of the Transplants Centre of the Regione Piemonte for providing the hCFs used in this work. This work was partially supported by Regione Piemonte CIPE 2007 Converging technology Project Metregen - Grant Agreement 20/07/2007.